\documentclass[vecphys]{svmult}

\usepackage{mathptmx}       
\usepackage{helvet}         
\usepackage{courier}        
\usepackage{type1cm}        
\usepackage{makeidx}         
\usepackage{graphicx}        
\usepackage{multicol}        
\usepackage[bottom]{footmisc}

\usepackage{color,soul}

\makeindex             

\usepackage[sort&compress,square,comma,numbers]{natbib}

\begin{document}

\title*{Brillouin optomechanics}

\author{Gaurav Bahl and Tal Carmon}

\institute{Gaurav Bahl \at University of Illinois at Urbana-Champaign, \email{bahl@illinois.edu}
\and Tal Carmon \at University of Michigan at Ann Arbor, \email{tcarmon@umich.edu}
}

\maketitle

\abstract{
In this chapter we introduce the concept of Brillouin optomechanics, a phonon-photon interaction process mediated by the electrostrictive force exerted by light on dielectrics and the photoelastic scattering of light from an acoustic wave.
We first provide a review of the phenomenon and continue with the first experiments where stimulated Brillouin optomechanical actuation was used in microdevices, and spontaneous Brillouin cooling was demonstrated.
}

\section{Introduction}
\label{sec:1}

Stimulated Brillouin scattering (SBS)\index{stimulated Brillouin scattering}\index{SBS} \cite{PhysRev.137.A1787,Yariv:1965ub} has been used since the 1960's as an acousto-optic gain mechanism for lasers \cite{PhysRevLett.12.592}. Subsequently, it has been employed as a tool for slow light \cite{PhysRevLett.94.153902}, for non-destructive characterization of materials \cite{rich:264,BIP:BIP360261002,Cheng:2006ju}, and in optical phase conjugation \cite{Zeldovich72} for holography. Brillouin lasers, due to their narrow linewidth, are also employed in ring laser gyroscopes \cite{ZARINETCHI:1991vw}. On the other hand, in fiber-based communication systems and high power fiber lasers, SBS is considered an undesirable mechanism that interferes with proper function \cite{SBS_Limits_Communications}. It is important to note that SBS is an optical nonlinearity common to all dielectrics in all states of matter, and is generally considered to be the strongest optical nonlinearity \cite{Boyd}.
In addition to optical fibers and bulk media SBS has been demonstrated in a variety of optical systems. Examples range over a variety of size scales, including nano-scale spheres \cite{PhysRevLett.90.255502}, fluid droplets \cite{Zhang:89}, photonic-crystal fibers \cite{Dainese:2006p1365}, and recently in mm-scale \cite{GrudininCaF2lasing,Lee:2012hn} and micron-scale resonators \cite{Tomes2009}.

\begin{figure}[tbp]
	\includegraphics[width=\columnwidth, clip=true, trim=0.5in 5.8in 1.2in 0.7in]{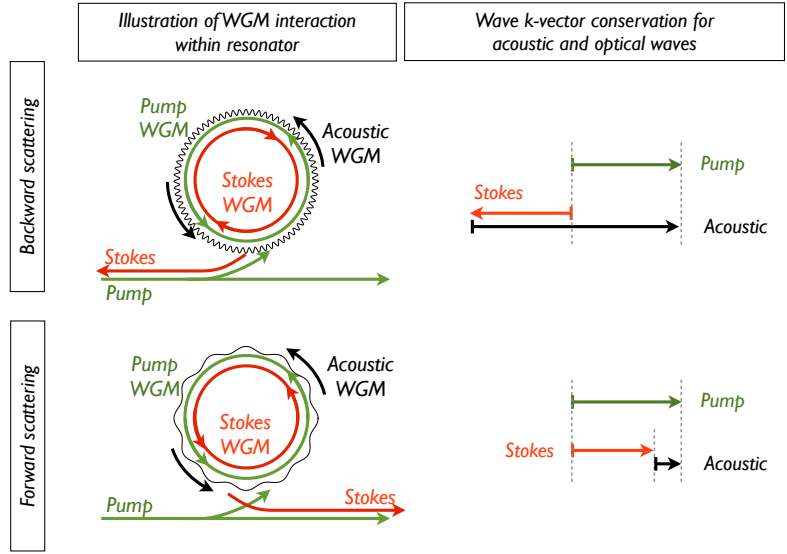}
	\caption{	
Momentum conservation dictates the frequencies of the acoustic modes that are coupled to optical modes via stimulated Brillouin scattering. 
Light evanescently couples into the resonator's optical WGMs from the tapered waveguide. Photons at the appropriate Stokes frequency are scattered in either the forward- or back-scattering direction.
	\textbf{Top:} In the case of back-scattering momentum conservation dictates that the acoustic wavevectors are long. The acoustic modes are therefore at high frequencies ($>$10 GHz). 
	\textbf{Bottom:} In forward scattering momentum conservation dictates that the acoustic modes are low frequencies ($<$1 GHz).
	}
	\label{fig:BC_1}       
\end{figure}

SBS is a process where light scatters from a sound wave in a material, resulting in a red-shifted (Stokes) laser and  amplification of the original sound wave. 
Phase-matching considerations require that the pump optical mode and the Stokes optical mode are separated by the precise energy and momentum of the acoustic phonons populating the sound wave.
Since light scattering can occur in either forward or backward direction relative to the pump, the momentum conservation constraint is a primary factor in determining the acoustic frequency. As shown in Fig.~\ref{fig:BC_1} back-scattered SBS is associated with large phonon momenta and therefore typically occurs with multi-GHz frequency acoustic waves \cite{GrudininCaF2lasing,Tomes2009,Lee:2012hn}. For instance, in fused silica using a pump wavelength of 1550~nm, back-scattered SBS creates an 11~GHz shifted Stokes laser relative to the pump. Forward scattered SBS, on the other hand, occurs with much lower acoustic frequencies ranging from the MHz to the low-GHz \cite{Bahl:2011cf,Savchenkov:11}. Forward Brillouin scattering has been experimentally observed previously in optical fibers \cite{Shelby:1985p1169}, photonic crystal fibers \cite{Dainese:2006p1365}, and recently in platforms where the mechanical as well as the optical waves are resonantly enhanced \cite{Bahl:2011cf,Savchenkov:11,Bahl_NComms2013}.

Resonators are convenient devices with which to investigate SBS since resonant enhancement enables very low lasing thresholds. However, in nano-scale resonators \cite{PhysRevLett.90.255502} this process suffers from very low optical finesse, while in larger resonators \cite{GrudininCaF2lasing,Tomes2009} the mechanical finesse was seen to be very low for back-scattered SBS due to high mechanical dissipation at 10-GHz rates. Given a system where optical modes and acoustic modes are simultaneously resonant, the Brillouin interaction is significantly enhanced.
It was later proposed \cite{MatskoSAWPRL} that forward-SBS, would allow access to acoustic modes at low frequencies where phonon lifetimes are longer, and therefore have high mechanical quality factors. 
Indeed, experimental studies of forward SBS in resonators have exhibited acoustic vibrations occurring over frequencies in the 10's-100's of MHz range in amorphous silica microspheres \cite{Bahl:2011cf}, silica microcapillaries \cite{Bahl_NComms2013}, and also in crystalline (LiTaO${}_3$ and MgF${}_2$) resonators \cite{Savchenkov:11}. These lower frequency vibrational modes were determined to have long phonon lifetimes, up to 2 orders of magnitude longer than the photon lifetimes \cite{Bahl:2011cf,BahlNP2012}. 
Such low phonon damping rates are needed for optomechanical cooling \cite{Grudinin:2010fe} as the rate of energy removal in the form of photons must exceed the rate of energy entering the acoustic mode from the bath.
Therefore, as we will show later, these results enabled reversal of the energy flow in the Brillouin process and demonstration of cooling \cite{BahlNP2012,PhysRevA.84.063806}.

Brillouin scattering thus joins other optomechanical actuation mechanisms in micro-devices, uniquely supporting vibration rates over a wide frequency range from 50~MHz to 12~GHz \cite{Bahl:2011cf,Bahl_NComms2013}, with the ability to cool mechanical whispering-gallery modes as well \cite{BahlNP2012}. Brillouin actuation acts on azimuthally circulating density waves in these devices, compared against actuation by radiation pressure on the device walls \cite{Carmon2005}. It is interesting to note that in the very same microresonator centrifugal radiation pressure can excite the sphere breathing mode \cite{Carmon2007} while at the same time Brillouin actuation can excite its mechanical whispering gallery modes \cite{Tomes2009,Bahl:2011cf,Bahl_NComms2013}. This simultaneous actuation was recently demonstrated shown in \cite{KewenFiO2013}.

\section{Brillouin optomechanics in resonators}
\label{sec:BC2}
\index{Brillouin optomechanics}

The Brillouin optomechanical process involves two optical modes that are separated in frequency space by the frequency of the acoustic mode of interest, and in momentum space by the momentum of the acoustic mode. These optical modes can be obtained by either designing the free spectral range (FSR) of the resonator, or by exploiting aperiodic spacing between high transverse-order optical modes in a resonator \cite{Savchenkov_StoppedLight,Carmon_StaticEnvelope,Bahl:2011cf}. Non-periodic spacing between resonances is particularly helpful in suppressing scattering in the Stokes direction when only anti-Stokes scattering is desired. This selective filtering capability was exploited to achieve spontaneous Brillouin cooling\index{Brillouin cooling} of the acoustic modes \cite{BahlNP2012,PhysRevA.84.063806} as we will discuss in the next section.

\subsection{Theoretical description}

We consider an optically stimulated traveling acoustic wave (i.e. sound wave) of frequency $\Omega_a$ traveling unidirectionally at the speed of sound at the equator of a whispering-gallery resonator. We assume this acoustic wave to be resonant, which happens when the circumference of the device is an integer multiple of the acoustic wavelength. This acoustic wave therefore forms an acoustic whispering-gallery mode (AWGM)\index{acoustic whispering gallery mode}. In order to describe the momentum associated with this acoustic mode, we use the integer propagation constant $M_a$. This integer is equal to the number of acoustic wavelengths present along the device equator and is therefore also known as the azimuthal mode order.
More formally, this integer represents the azimuthal propagation of the AWGM along the device as $e^{j(M \cdot \phi -\omega \cdot t)}$ where $\phi$ is the azimuthal position and $\omega$ is the wave frequency.

\begin{figure}[tbp]
	\includegraphics[width=\columnwidth, clip=true, trim=2in 4.5in 2in 0.2in]{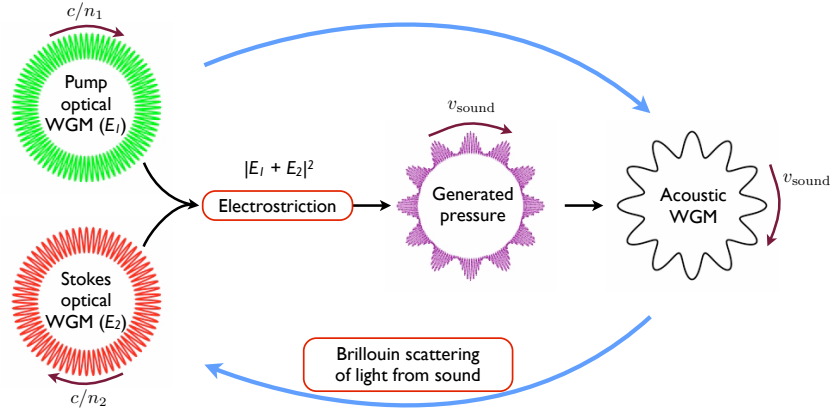}
	\caption{	
Here we illustrate the optomechanical interaction within a resonator in the case of forward-SBS. The acoustic WGM writes a traveling photoelastic grating that scatters light from the pump OWGM to the Stokes OWGM, i.e. Brillouin scattering. Simultaneously, the electrostrictive interference pattern generated by the two OWGMs amplifies the acoustic WGM. In this manner a positive feedback is generated. Note that the two OWGMs travel at the speed of light ($c/n_1$ and $c/n_2$ where $n_{1,2}$ are the refractive indices), and the AWGM travels at the speed of sound ($v_{sound}$).
	}
	\label{fig:BC_Loop}    
\end{figure}

This traveling mechanical mode photo-elastically writes an optical grating that is capable of scattering light present in any optical whispering-gallery mode (OWGM) that has good spatial overlap with this AWGM.\index{photoelastic scattering}\index{Brillouin scattering}
As a result of this scattering, pump light is scattered to red-shifted (Stokes) frequencies as shown in Fig.~\ref{fig:BC_1} for both forward and backward scattering directions. The relationship between the pump (subscript p), Stokes (subscript S), and acoustic (subscript a) wave frequencies is given by
\begin{equation}
	\omega_S = \omega_p - \Omega_a
	\label{Eq1}
\end{equation}
which is essentially the energy conservation relationship. Note that momentum conservation between photons and phonons must also be accounted, as we describe below.
Simultaneously, the pump light and newly scattered Stokes light generate a beat note, i.e. a spatio-temporal interference pattern within the resonator (Fig.~\ref{fig:BC_Loop}, purple ``Generated pressure''). When we evaluate the electrostrictive force generated by this interference pattern, we see that the optically induced stress appears precisely at the spatial and temporal frequency of the original acoustic wave, and travels at the speed of sound. As a result of this positive feedback, this process described by photoelastic scattering and electrostriction becomes self-sustaining. The acoustic wave is amplified due to this feedback, and a Stokes laser is generated. The complete optomechanical interaction between the two OWGMs and the AWGM is illustrated for the case for forward-SBS in Fig.~\ref{fig:BC_Loop}.

We can derive this intuitive phase-matching condition (Eqn.~\ref{Eq1}) analytically by solving the coupled wave equations for mechanical and optical waves to reveal a synchronous solution \cite{Boyd}. 
A detailed solution of the coupled acousto-optic interaction is presented in the supplementary information of ref. \cite{BahlNP2012}. This solution reveals that the propagation constants of the pump OWGM, $M_p$, must equal the sum of the propagation constants of the resulting Stokes and acoustic modes, as given by
\begin{equation}
	M_p = M_S + M_a  ~.
\end{equation}
Since one pump photon is converted into a photon and a phonon whose combined momentum must equal the pump photon's momentum, this wavevector relation is associated with momentum conservation.

This momentum relationship is graphically represented in Fig.~\ref{fig:BC_1} for both forward and backward scattering. As can be seen, the acoustic momentum vector in the case of backscattering is roughly double the length of the optical vectors. As a result the acoustic wavelength in the back-scattering case is half that of the optical wavelength used, and the acoustic frequencies are very high. Acoustic frequencies in the 10 GHz - 17 GHz regime have been observed in various glass and crystalline resonators \cite{GrudininCaF2lasing, Tomes2009}.
On the other hand, since the acoustic wavevector is very short in the forward-scattering case, the acoustic frequencies are lower and the acoustic wavelengths are much longer. This prediction \cite{MatskoSAWPRL} was verified experimentally with silica microsphere resonators \cite{Bahl:2011cf}, which we describe in the next subsection.

In the case of back-scattering, the phonon lifetimes are low, as a result of which the propagation distance is comparable with the device circumference, implying a low mechanical finesse. This lack of acoustic finesse is a concern even for resonators in the 100 ${\mu}m$ size scale. On the other hand, it is known that phonon dissipation from the material scales inversely as the square of frequency \cite{Boyd}. This reduced dissipation increases the phonon lifetimes significantly at lower acoustic frequencies. The implication is that a reduction in acoustic frequency from the 11\,000 MHz (typical of backward scattering in silica) to the 50 MHz regime (typical in forward scattering SBS) may potentially provide a 50\,000-fold reduction in mechanical dissipation. The result is that phonons survive long enough to circumnavigate the WGR many times, or in other words the phonon finesse is $\gg 1$. The stimulated acoustic wave now becomes a high-finesse mechanical resonance.

As mentioned above, a key requirement of achieving SBS-based optomechanical interaction within a microresonator is the availability of two optical modes with nearly identical optical frequencies, but with propagation constants differing by $M_a$.
However, the frequency separation between optical modes having adjacent $M$-numbers within one mode family is defined by the resonator FSR. Since the FSR of the optical modes of a silica resonator of 100 micron diameter is $>$300 GHz, the two required optical modes cannot belong to the same optical mode family, for either the forward- or back-scattering cases. This leads to a significant experimental challenge with obtaining the requisite phase match.
On the other hand, optical modes with high-orders in the transverse plane exhibit different effective refractive indices. As a result, the optical wavevectors may differ significantly between two optical modes that have nearly identical frequency. The existence of such high order modes of different azimuthal wavelength but almost the same frequency was indicated by standing interference patterns experimentally generated in \cite{Savchenkov_StoppedLight,Carmon_StaticEnvelope}. Use of such modes in achieving phase match for SBS was discussed in \cite{MatskoSAWPRL}.

\begin{figure}[tbp]
\centering
	\includegraphics[width=\columnwidth, clip=true, trim=0.4in 5in 1.2in 0.8in]{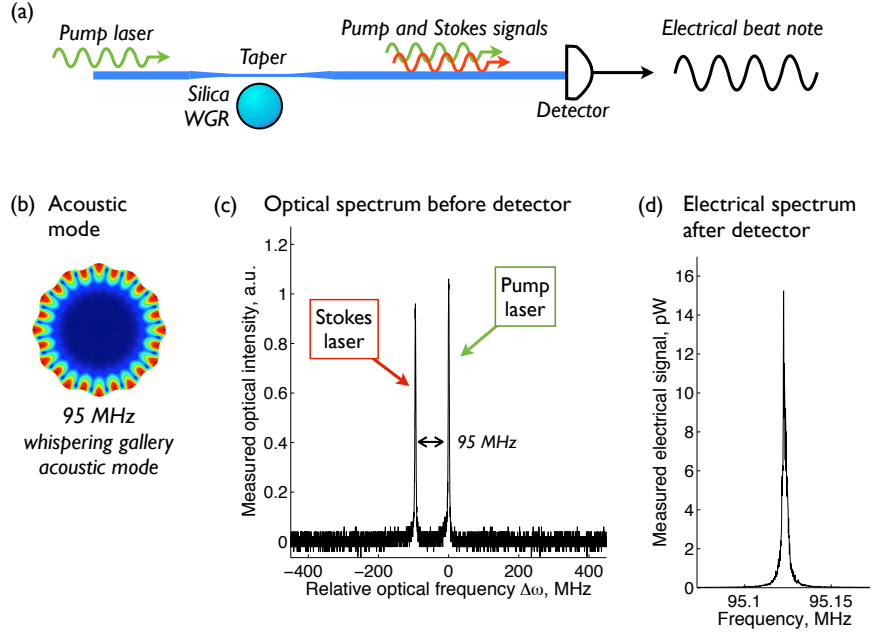} 
	\caption{	
	\textbf{(a)} Experimental setup
	\textbf{(b)} Finite element calculation for the acoustic whispering-gallery mode in this experiment.
	\textbf{(c)} Pump and stokes laser at output
	\textbf{(d)} Electrical beat signal provides acoustic signature
	\label{fig:BC_Heat}    }  
\end{figure}

\subsection{Experimental implementation}

We perform the experiment \cite{Bahl:2011cf} on a silica microsphere resonator \cite{GorodetskyMicrospheres} having a typical optical $Q > 10^8$. A tapered optical fiber is employed to evanescently couple light into the pump OWGM \cite{Cai:01, Spillane:2003p1318, Knight:1997p1319}, and a telecom-wavelength laser at 1550 nm is used as the source as shown in Fig.~\ref{fig:BC_Heat}a.

When we tune the pump laser to a resonance of the microsphere where the requisite phase match exists, pump photons are forward-scattered to the Stokes OWGM upon interaction with the thermal phonons in the AWGM. This scattered light also evanescently couples out to the tapered optical fiber and propagates to a photodetector at the opposite end of the fiber (Fig.~\ref{fig:BC_Heat}a). The partially transmitted pump and the newly generated Stokes light (Fig.~\ref{fig:BC_Heat}c) create a temporal beat note at the detector (Fig.~\ref{fig:BC_Heat}d), which is at the frequency of the acoustic mode (Fig.~\ref{fig:BC_Heat}b). This is a direct measure of the acoustic mode, and therefore the phonon profile in the resonator. We also employ optical spectrum analyzers to observe the optical signals in order to verify the SBS phenomenon (Fig.~\ref{fig:BC_Heat}c). No modulation of the laser nor any feedback control is needed to create this stimulated vibration.

\begin{figure}[btp]
	\includegraphics[width=\columnwidth, clip=true, trim=2.8in 3.4in 2.6in 0.4in]{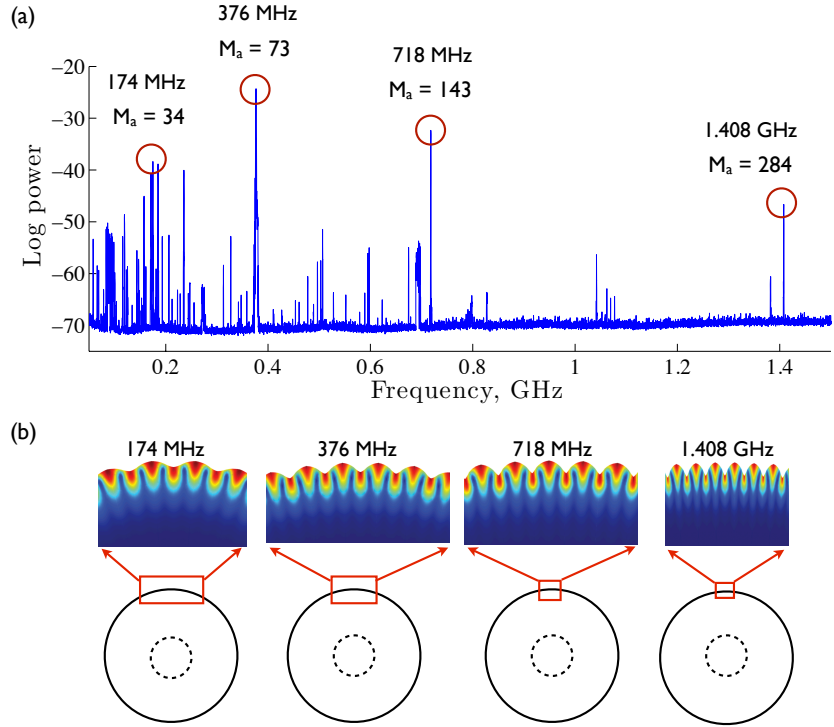}
	\caption{	
\textbf{(a)} Many acoustic WGMs ranging from 50 MHz - 1.4 GHz are observed when the pump laser is scanned slowly from 1,520 - 1,570 nm. Some strong modes are identified here along with their propagation constant, $M_a$. \textbf{(b)} Equatorial cross-section of the finite-element calculated mode shapes for the modes identified in (a).
	}
	\label{fig:BC_Freq}       \end{figure}

Indeed, for a given resonator, hundreds of instances of phase match can be identified experimentally. In \cite{Bahl:2011cf} we showed that mechanical frequencies ranging from 58 MHz - 1.4 GHz (Fig.~\ref{fig:BC_Freq}) could be generated on a single device when the laser was tuned over 1520nm - 1570nm. The lowest measured threshold power was 22.5 ${\mu}W$ for forward Brillouin lasing at 175 MHz. Mechanical mode quality factors are estimated by observing the subthreshold linewidth as we describe later in the subsection on Brillouin cooling.

\section{Spontaneous Brillouin cooling}
\label{sec:BC3}
\index{Brillouin cooling}

Electrostrictive Brillouin scattering is often only thought of as an amplification process \cite{PhysRevLett.12.592,Boyd,Shelby:1985p1169,Dainese:2006p1365,GrudininCaF2lasing,Tomes2009,Bahl:2011cf,Bahl_NComms2013} since it readily appears in most optical systems when a threshold power is surpassed. On the other hand it is a known fact that spontaneous anti-Stokes Brillouin scattering also occurs \cite{Boyd}, which based on energy conservation arguments should lead to annihilation of phonons and therefore cooling of the acoustic mode.
As we see in the rest of this book, bolometric \cite{Mertz_Photothermal,Metzger:2004p1357,PhysRevLett.101.133903} and ponderomotive \cite{Arcizet:2006p1092, Gigan:2006p1091, Kleckner:2006p1082, Thompson:2008p1083, Riviere:2011cj, Chan_GroundState_2011} forces are now widely used for laser cooling of the mechanical modes of microdevices. The result we present here demonstrates this cooling capability with the Brillouin optomechanical process.

\subsection{Cooling theory}

It had been suggested that cooling would be possible in multi-resonance systems, such as those discussed in refs. \cite{Grudinin:2010fe,GrudininCaF2lasing,Tomes2009,Bahl:2011cf,Bahl_NComms2013}, provided that the phonon damping rate is lower than the damping rate of photons \cite{Grudinin:2010fe}. This is required so that the rate of thermal energy removal from the acoustic mode in the form of photons is greater than the rate at which energy re-enters the acoustic mode. However, this regime is not accessible in the SBS back-scattering experiments \cite{PhysRevLett.12.592,Boyd,Shelby:1985p1169,GrudininCaF2lasing,Tomes2009} since phonons at $>$10GHz acoustic frequencies are generally associated with high mechanical damping \cite{Boyd}. 
On the other hand, forward Brillouin scattering \cite{Shelby:1985p1169} in microcavities \cite{Bahl:2011cf} provides a path to optomechanically couple 10-100 MHz range AWGMs where the mechanical dissipation is far lower, as measured through the mechanical quality factor.

A second challenge remains, which is to eliminate Stokes scattering. 
In such light-sound interactions incident photons are scattered to both redder (Stokes) and bluer (anti-Stokes) frequencies, resulting in heating and cooling as dictated by energy conservation. It was thought that this cooling-heating balance is always tilted towards heating (i.e. Stokes scattering) as governed by Planck's distribution \cite{Kittel}. This is indeed true in bulk media where all photons are almost equally transmitted. 
Furthermore, Stokes scattered light applies a positive feedback resulting in lasing through stimulated Brillouin scattering (SBS) that creates a large influx of phonons into the acoustic mode.
We showed that this heating-cooling balance can be tilted towards cooling by exploiting the aperiodic nature of optical resonances in a resonator.
In particular, rejection of Stokes scattering can be achieved by engineering the optical density of states in the system to eliminate any available phase-matched Stokes optical mode. Such filtering is challenging to do in bulk materials since relatively low (i.e. MHz regime) acoustic frequencies separate the optical modes, and sharp filtering transitions are not easily available. On the other hand, ultra-high-Q resonators readily exhibit optical modes with mode linewidths in the MHz regime, allowing for selective suppression of Stokes scattering.

We present a classical analysis in \cite{BahlNP2012}, showing that the cooling ratio is improved by having higher acoustic and optical quality factors and lower acoustical frequencies. Our quantum analysis in \cite{Tomes_coolingTheory} provides details on the feasibility of using this system to reach ground-state \cite{Chan_GroundState_2011,Riviere:2011cj,Teufel:2011jg}.

The concept of using anti-Stokes scattering to cool a material was originally proposed by Pringsheim \cite{Pringsheim1929}. However, concerns regarding the thermodynamic impossibility of such a process were later raised as the second law of thermodynamics is seemingly violated. The concern was eventually resolved by Landau \cite{Landau1946} with the explanation that the entropy of scattered light is increased through decoherence in phase, frequency, and directionality. In our experiment, the linewidth broadening of the scattered light is responsible for carrying the excess entropy as expected.

\subsection{Cooling experiment}

We again use a silica microsphere resonator, which supports the three modes that participate in the Brillouin cooling process -- two OWGMs (pump and anti-Stokes) and one AWGM. Since this is a forward-scattering interaction these three modes modes circulate in unison with considerable overlap.  As mentioned previously, photoelastic scattering induced by density change scatters light to the anti-Stokes mode. Cooling of the acoustic mode occurs by means of the electrostrictive pressure generated by the light in the pump OWGM and the anti-Stokes OWGM, which acts to attenuate the themal mechanical motion of the phase matched AWGM.

\begin{figure}[tbp]
\centering
	\includegraphics[width=\columnwidth, clip=true, trim=0.8in 5.5in 1.2in 1.2in]{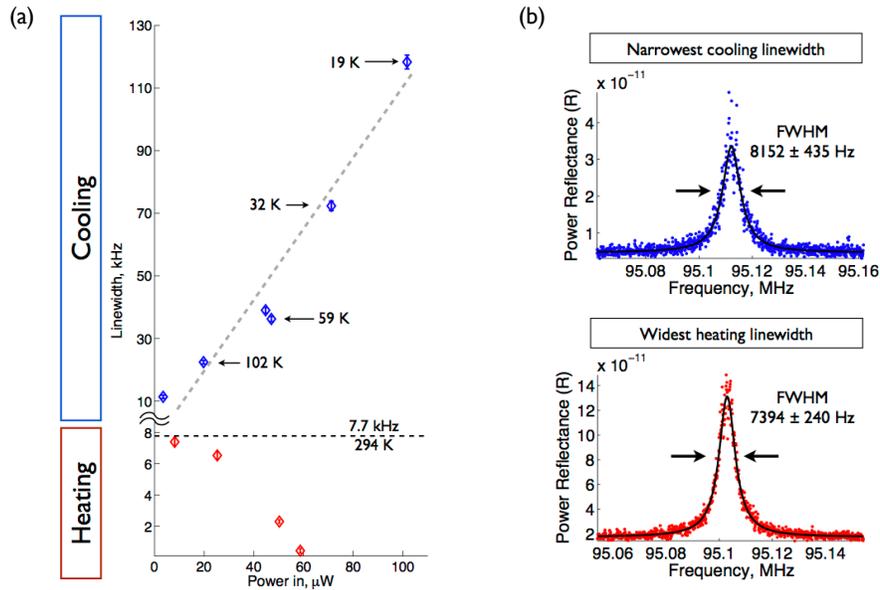}
	\caption{	
	\textbf{(a)} Acoustic linewidth data as a function of input optical power during heating and cooling experiments.
	\textbf{(b)} The 294\,K calibration is performed by measuring the narrowest cooling linewidth and the widest heating linewidth. The trends are extrapolated back to zero input optical power in order to obtain unperturbed intrinsic linewidth.
	\label{fig:BC_Cool}    }   
\end{figure}

The pump laser is positioned at the lower frequency optical mode. Light scatters in the anti-Stokes direction into the higher frequency optical mode. In the process, phonons are annihilated from the acoustic mode. Brillouin cooling is experimentally observed as the broadening of the beat-note signal between pump light and anti-Stokes lines, as a function of increasing pump power.

We employ acoustic mode linewidth as a measurement of cooling as was also shown in \cite{Metzger:2004p1357,Arcizet:2006p1092}. The acoustic mode's effective temperature is inversely proportional to linewidth.
In our experiment, we measured a 7.7~kHz sub-threshold linewidth for the 95 MHz acoustic mode at 294 K (Fig.~\ref{fig:BC_Cool}b). The linewidth increased to 118 kHz when the pump power was increased to 100 $\mu$W, indicating a cooling ratio of 15 (Fig.~\ref{fig:BC_Cool}a). As a result, the effective mode temperature at the maximum cooling point was calculated as 19~K.
To ensure that the 7.7 kHz acoustic linewidth corresponds to the Brownian vibration, the Stokes (heating) experiment was performed with the pump laser positioned on the higher frequency optical resonance, and the acoustic linewidth was measured as a function of input power. The 7.7~kHz acoustic linewidth was verified by extrapolating both the heating and cooling linewidth trends to zero input optical power (Fig.~\ref{fig:BC_Cool}a,b).
The end result in the Brillouin cooling experiment is a net annihilation of phonons \cite{EpsteinBook,Chan_GroundState_2011, Teufel:2011jg,Riviere:2011cj, Arcizet:2006p1092, Gigan:2006p1091, Kleckner:2006p1082,Thompson:2008p1083,Metzger:2004p1357} in a multi-resonance device \cite{Grudinin:2010fe,GrudininCaF2lasing,Tomes2009,Bahl:2011cf}. However, it is important to note that the pump does not need to be detuned with respect to the optical resonance, which greatly simplifies the experimental configuration.

Broadly speaking, Brillouin scattering belongs to a family of material-level scattering processes, including Raman scattering and Rayleigh scattering. The ability to remove phonons from a material by means of Brillouin scattering raises the question of whether similar cooling experiments would be possible with optical phonons (Raman scattering) \cite{Kang:2009p1366} in spite of the significantly higher phonon frequencies involved. Raman cooling is attractive since it can provide a considerably higher quantum cooling efficiency.

\begin{acknowledgement}
This work was supported by the Defense Advanced Research Projects Agency (DARPA) Optical Radiation Cooling and Heating in Integrated Devices (ORCHID) program through a grant from the Air Force Office of Scientific Research (AFOSR). Additional thanks go to the University of Illinois Mechanical Science and Engineering startup grant. The authors would also like to acknowledge the contributions of Matthew Tomes, John Zehnpfennig, and Kyu Hyun Kim.
\end{acknowledgement}


\end{document}